\documentclass{PoS}

\title{A radiative seesaw model with GeV singlet-doublet fermion and TeV triplet scalar dark matter}

\ShortTitle{A radiative seesaw model with singlet-doublet fermion and triplet scalar dark matter}

\author{Juri Fiaschi, \speaker{Michael Klasen}\thanks{Work supported by the BMBF under contract
  05H18PMCC1 and the DFG through the Research Training Group 2149 ``Strong and weak
  interactions -- from hadrons to dark matter''.}\\
  Institut f\"ur Theoretische Physik,
  Westf\"alische Wilhelms-Universit\"at M\"unster,
  Wilhelm-Klemm-Stra\ss{}e 9,
  48149 M\"unster, Germany\\
  E-mail: \email{fiaschi@uni-muenster.de}, \email{michael.klasen@uni-muenster.de}}

\author{Simon May\\
  Max-Planck-Institut f\"ur Astrophysik,
  Karl-Schwarzschild-Stra\ss{}e 1,
  85741 Garching, Germany\\
  E-mail: \email{simon.may@mpa-garching.mpg.de}}

\abstract{By extending the Standard Model with singlet-doublet fermions and
  triplet scalars, all odd under a new $Z_2$ symmetry, we introduce a radiative
  seesaw model that can simultaneously account for dark matter, explain the
  existence of neutrino masses and allow for gauge coupling unification. We
  explore the viable parameter space of the model after imposing collider,
  Higgs mass, dark matter, neutrino mass and lepton flavour violation
  constraints. We find that dark matter in this model is fermionic for
  masses below about 1 TeV and scalar above and observe a high degree of
  complementarity between direct detection and lepton flavour violation
  experiments, which should soon allow to fully probe the fermionic dark
  matter sector and at least partially the scalar dark matter sector.}

\FullConference{
European Physical Society Conference on High Energy Physics - EPS-HEP2019 -\\
			10-17 July, 2019\\
			Ghent, Belgium}

\usepackage{booktabs}
\usepackage{amsmath}

\newcommand{\beq}{\begin{equation}}
\newcommand{\eeq}{\end{equation}}
\newcommand{\bea}{\begin{eqnarray}}
\newcommand{\eea}{\end{eqnarray}}

\def\lr{\left( }
\def\rr{\right) }

\begin{document}

\section{Motivation}

Evidence from many different length scales for dark matter (DM) and precision
measurements of its relic abundance in the Universe, together with small, but
non-zero neutrino masses and an unnaturally light Higgs boson are clear
indications that the Standard Model (SM) of particle physics is incomplete.
While supersymmetry has long been favoured as a global
solution to these and other puzzles, supersymmetric particles have so far not
been discovered despite intense searches at the LHC and in direct
DM detection experiments \cite{Klasen:2015uma}.

A less encompassing, rather bottom-up approach to these problems consists in
extending the SM minimally, e.g.\ by additional Higgs multiplets $\phi_i$,
which would, however, not acquire a vacuum expectation value, and/or additional
right-handed neutrinos $\nu_R$, that are also employed in different types of
seesaw mechanisms to generate the SM neutrino masses. When imposing an additional
$Z_2$ symmetry, under which both $\phi_i$ and $\nu_R$ are odd, it is possible
to both render the lightest inert particle into a stable DM candidate, and to
avoid tree-level couplings of single $Z_2$-odd particles to the those of the SM.

\section{Radiative seesaw models}

Neutrino masses are then generated radiatively, and the corresponding one-loop
models have recently been classified \cite{Restrepo:2013aga}. For several
among them, we have already performed detailed phenomenological studies in the
past, e.g.\ for doublet scalar DM with singlet fermion coannihilations \cite{Klasen:2013jpa},
singlet scalar and singlet-doublet fermion DM \cite{Klasen:2016vgl}
and singlet-doublet scalar and singlet-doublet fermion DM \cite{Esch:2018ccs}.
We have also studied inert scalar DM with electroweak one-loop corrections
\cite{Klasen:2013btp}, singlet fermion DM interacting with a new singlet scalar
\cite{Esch:2013rta}, fermionic DM in the freeze-in mechanism \cite{Klasen:2013ypa},
two-component DM \cite{Esch:2014jpa} and scalar DM in the $B-L$ model
\cite{Klasen:2016qux}.

Here we present a recent study of a model with singlet-doublet fermions
and triplet scalars \cite{Fiaschi:2018rky}, dubbed T1-3-B with $\alpha=0$ in the
classification scheme cited above. The new fields and their quantum numbers in
this model are shown in Tab.\ \ref{tab:1}, and the interaction Lagrangian is
given by
\begin{table}[b]
	\centering
 	\caption{New fields and their quantum numbers in model T1-3-B with $\alpha = 0$.}
	\label{tab:1}
	\medskip
	\begin{tabular}{c c c c c c c c}
		\toprule
		Field & Generations & Spin & Lorentz rep. & $SU(3)_C$ & $SU(2)_L$ & $U(1)_Y$ & $\mathbb{Z}_2$\\
		\midrule
		$\Psi$ & 1 & $\frac{1}{2}$ & $(\frac{1}{2}, 0)$ & $\mathbf{1}$ & $\mathbf{1}$ & $0$ & $-1$ \\
		$\psi$ & 1 & $\frac{1}{2}$ & $(\frac{1}{2}, 0)$ & $\mathbf{1}$ & $\mathbf{2}$ & $-1$ & $-1$ \\
		$\psi'$ & 1 & $\frac{1}{2}$ & $(\frac{1}{2}, 0)$ & $\mathbf{1}$ & $\mathbf{2}$ & $1$ & $-1$ \\
		$\phi_i$ & 2 & $0$ & $(0, 0)$ & $\mathbf{1}$ & $\mathbf{3}$ & $0$ & $-1$ \\
		\bottomrule
	\end{tabular}
\end{table}
\bea
 \mathcal{L} &=& 
 - \frac{1}{2} (M_\phi^2)^{ij} Tr(\phi_i \phi_j)
 - \lr \frac{1}{2} M_\Psi \Psi \Psi + {\rm h.c.} \rr
 - \lr M_{\psi\psi'} \psi \psi' + {\rm h.c.} \rr \nonumber\\
 &&
 - (\lambda_1)^{ij} (H^\dagger H) Tr(\phi_i \phi_j)
 - (\lambda_3)^{ijkm} Tr(\phi_i \phi_j \phi_k \phi_m) \nonumber\\
 &&
 - \lr \lambda_4 (H^\dagger \psi') \Psi + {\rm h.c.} \rr
 \!-\! \lr \lambda_5 (H \psi) \Psi + {\rm h.c.} \rr
 \!-\! \lr { (\lambda_6)^{ij}} L_i \phi_j \psi' + {\rm h.c.} \rr.
\eea
After electroweak symmetry breaking, the neutral fermions acquire
the mass terms
\begin{equation}
 \mathcal{L}_{{\rm f},0} = -\frac{1}{2} M_\Psi \Psi \Psi - M_{\psi\psi'} \psi^0 \psi'^0
 -\frac{\lambda_4 v}{\sqrt{2}} \psi'^0 \Psi - \frac{\lambda_5 v}{\sqrt{2}} \psi^0 \Psi
 + {\rm h.c.},\nonumber
\end{equation}
which leads to the singlet-doublet fermion mass matrix and
corresponding eigenstates
\begin{equation}
 M_{{\rm f},0} = \begin{pmatrix} M_\Psi & \frac{\lambda_5 v}{\sqrt{2}} & \frac{\lambda_4 v}{\sqrt{2}} \\ \frac{\lambda_5 v}{\sqrt{2}} & 0 & M_{\psi\psi'} \\ \frac{\lambda_4 v}{\sqrt{2}} & M_{\psi\psi'} & 0\end{pmatrix}
\qquad {\rm with}\qquad
 \chi^0 = U_\chi \begin{pmatrix}\Psi^0 \\ \psi^0 \\ \psi'^0\end{pmatrix}.
\end{equation}
Two generations ($n_s=2$) of triplet scalars $\phi_i$ are required for two non-zero
SM neutrino masses. They obtain the mass matrices
\begin{equation}
 M_{\phi^0}^2 = M_{\phi^\pm}^2 = M_\phi^2 + \lambda_1 v^2.
\qquad {\rm with}\qquad
 \eta^{0,\pm} = O_\eta \begin{pmatrix} \phi_1^{0,\pm} \\ \phi_2^{0,\pm} \end{pmatrix}.
\end{equation}
Note that the charged scalars are slightly heavier than
their neutral counterparts due to one-loop electroweak diagrams
by about \cite{Cirelli:2005uq}
\begin{equation}
 \Delta m_{\eta_i} = m_{\eta_i^{\pm}} - m_{\eta_i^0} = 166\ {\rm MeV}.
\end{equation}


When one decouples the scalars by setting
\begin{equation}
 (M_\phi^2)^{11} = 
 (M_\phi^2)^{22} = (1000\ {\rm TeV})^2,\
 (M_\phi^2)^{12} = 0,\
 \lambda_1 = \lambda_3 = \lambda_6 = 0
\end{equation}
while keeping the fermion masses light and couplings non-zero as in
\beq
 M_\Psi = 200\ {\rm GeV},\
 M_{\psi\psi'} = 300\ {\rm GeV},\
 \lambda_5 = 0.36,
\eeq
we reproduce the relic density and direct detection cross sections
predicted in the literature \cite{Cohen:2011ec,Cheung:2013dua}.
In our study, we update, however, the
Higgs boson mass and the nuclear form factors. As had been noted
before, blind spots of spin-independent (SI) and spin-dependent (SD)
direct detection can appear depending on the value of $\lambda_4$.
With the new results of the XENON1T experiment, the mass limits
increase to $M_\Psi \simeq M_{\psi\psi'} > 200$ GeV ... 1 TeV.


When one decouples instead the fermions by setting
\begin{equation}
 M_\Psi = M_{\psi\psi'} = 1000\ {\rm TeV}
 ,\
 \lambda_4 = \lambda_5 = \lambda_6 = 0
\end{equation}
and also the unimportant scalar self coupling $\lambda_3 = 0$,
we find that for one generation $m_\eta\simeq2$ TeV except for
large Higgs couplings $\lambda_1$, where $m_\eta$ must even be
in the multi-TeV region. In this case we found that we had to
correct a result in the literature by a normalisation factor
of two in the squared neutral mass \cite{Araki:2011hm}.

\section{Radiative neutrino masses}

\begin{figure}[b!]
  \centering
  \includegraphics[width=0.4\textwidth]{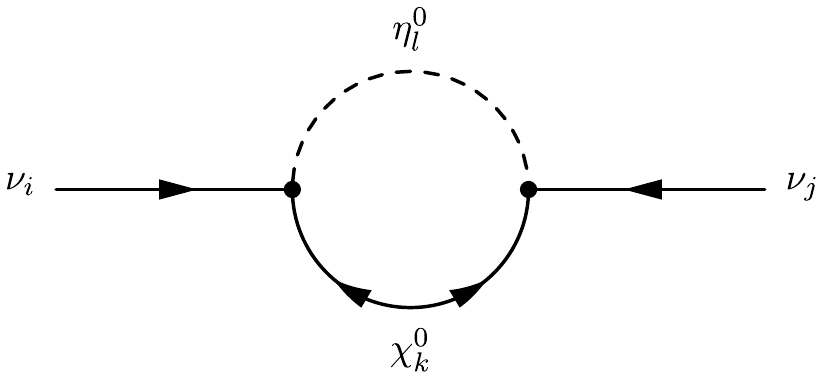}
 \caption{Feynman diagram for the radiative generation of neutrino masses at
		the one-loop level ($k \in \{1, \dots, n_{\text{f}}\}$, $l \in \{1, \dots, n_{\text{s}}\}$). \label{fig:1}}
\end{figure}

After electroweak symmetry breaking, the SM neutrino masses are generated
at one loop through the Feynman diagram shown in Fig.\ \ref{fig:1}.
The corresponding neutrino mass matrix is given by
\bea
 (M_\nu)_{ij}
 &=& \frac{1}{32\pi^2} \sum_{l=1}^{n_{\text{s}}} { \lambda_6^{im} \lambda_6^{jn}} (O_\eta)_{ln} (O_\eta)_{lm}
 \sum_{k=1}^{n_{\text{f}}} {(U_\chi)^*_{k3}}^2 \frac{m_{\chi_k^0}^3}{m_{\eta^0_l}^2 - m_{\chi_k^0}^2} \ln\lr \frac{m_{\chi_k^0}^2}{m_{\eta^0_l}^2} \rr.
 \label{eq:neutrino_masses}
\eea
Since there are no neutrino masses at tree level, the ultraviolet divergences cancel
as they must. The mass matrix is diagonalised to $D_\nu$ with the PMNS matrix $U_\nu$.
When expanded in $\lambda_{4,5}\ll1$, it simplifies to
\begin{equation}
 M_\nu \approx 100\ {\rm meV}\ \frac{M_\Psi}{1\ {\rm TeV}}\
 \lr \frac{{ \lambda_6^{ij}} \lambda_{4,5}}{10^{-5}} \rr^2,
\end{equation}
showing that the singlet fermion mass is of order 1 TeV for couplings
$\lambda_{4,5,6}$ of about $10^{-2}$ or slightly below.
It is useful to diagonalise the neutrino mass matrix without the
important coupling $\lambda_6$,
\begin{equation}
 M_\nu = \lambda_6^T M \lambda_6 = \lambda_6^T U_M^T D_M U_M \lambda_6\nonumber
\end{equation}
which allows to express the latter in the so-called Casas-Ibarra
parametrisation \cite{Casas:2001sr}
\begin{equation}
 \lambda_6 = U_M^T D_M^{-\frac{1}{2}} R D_\nu^{\frac{1}{2}} U^\dagger_\nu\nonumber
 \label{eq:casas_ibarra}
\end{equation}
with an arbitrary rotation matrix
\begin{equation}
 R =
 D_M^{\frac{1}{2}} U_M \lambda_6 U_\nu D_\nu^{-\frac{1}{2}} \ =\ 
 \begin{pmatrix} 0 & \cos(\theta) & \sin(\theta) \\ 0 & -\sin(\theta) & \cos(\theta)\end{pmatrix}.
 \nonumber
\end{equation}
We can then directly impose the SM neutrino mass difference
and mixing constraints and scan the free parameters of the
model over the ranges $\theta\in[0;2\pi]$, $|\lambda_{1,4,5}|
\in[10^{-6};1]$, $\lambda_{4}>0$, and $M_\phi$, $M_\Psi$,
$M_{\psi\psi'}\in[10$ GeV$;10\,000$ GeV$]$. In addition, we
impose direct experimental constraints from LEP on
$m_{\chi^0,\eta^0}>m_Z/2$, $m_{\psi^-,\psi^{\prime +},\eta^\pm_i}>102$ GeV,
from the LHC on the Higgs boson mass $m_H=125\pm2.5$ GeV, and from Planck on
the DM relic density $\Omega_c^{\rm obs} h^2=0.120\pm0.001$.

\section{Fermion DM}

\begin{figure}
 \includegraphics[width=0.49\textwidth]{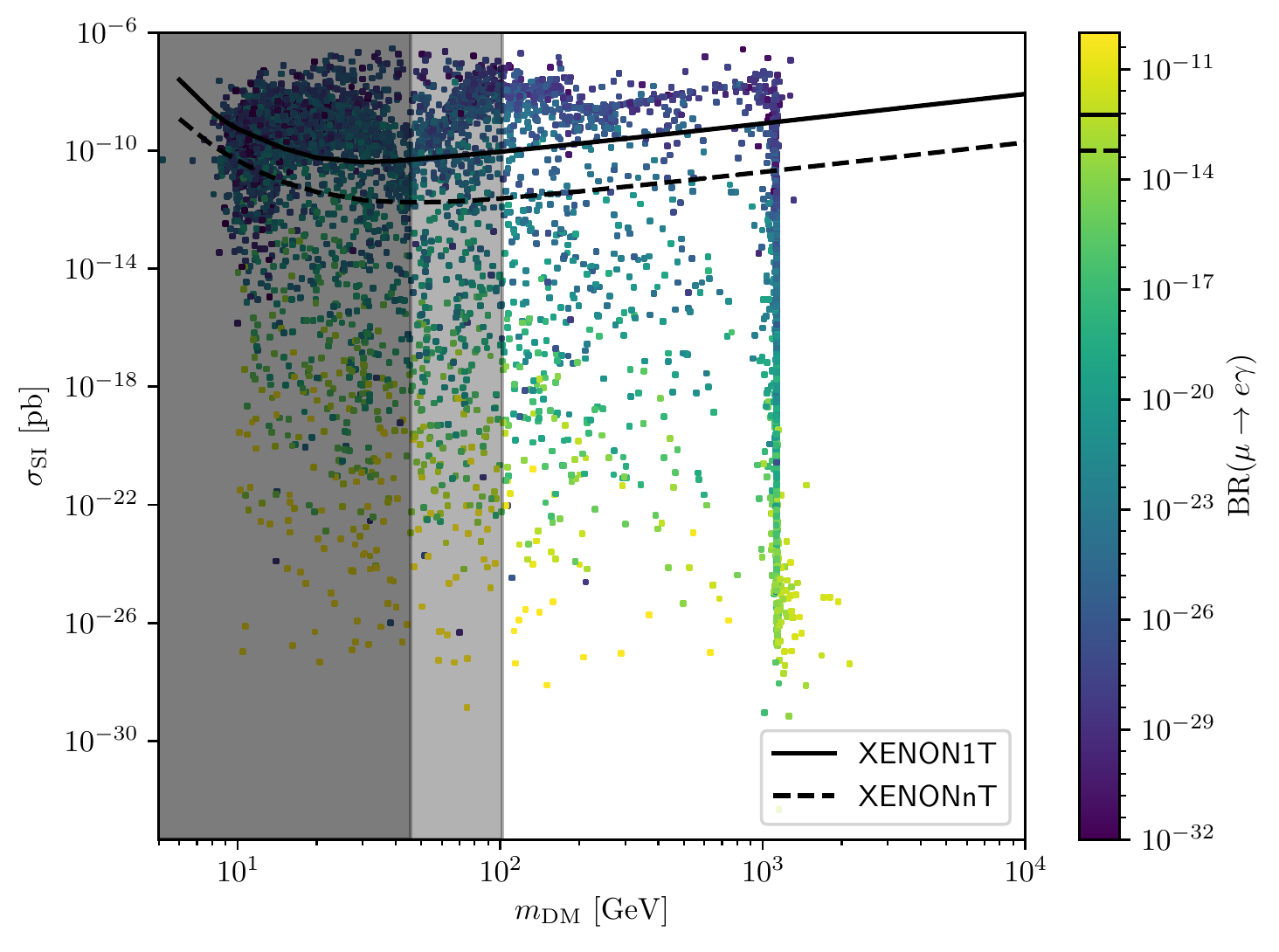}
 \includegraphics[width=0.49\textwidth]{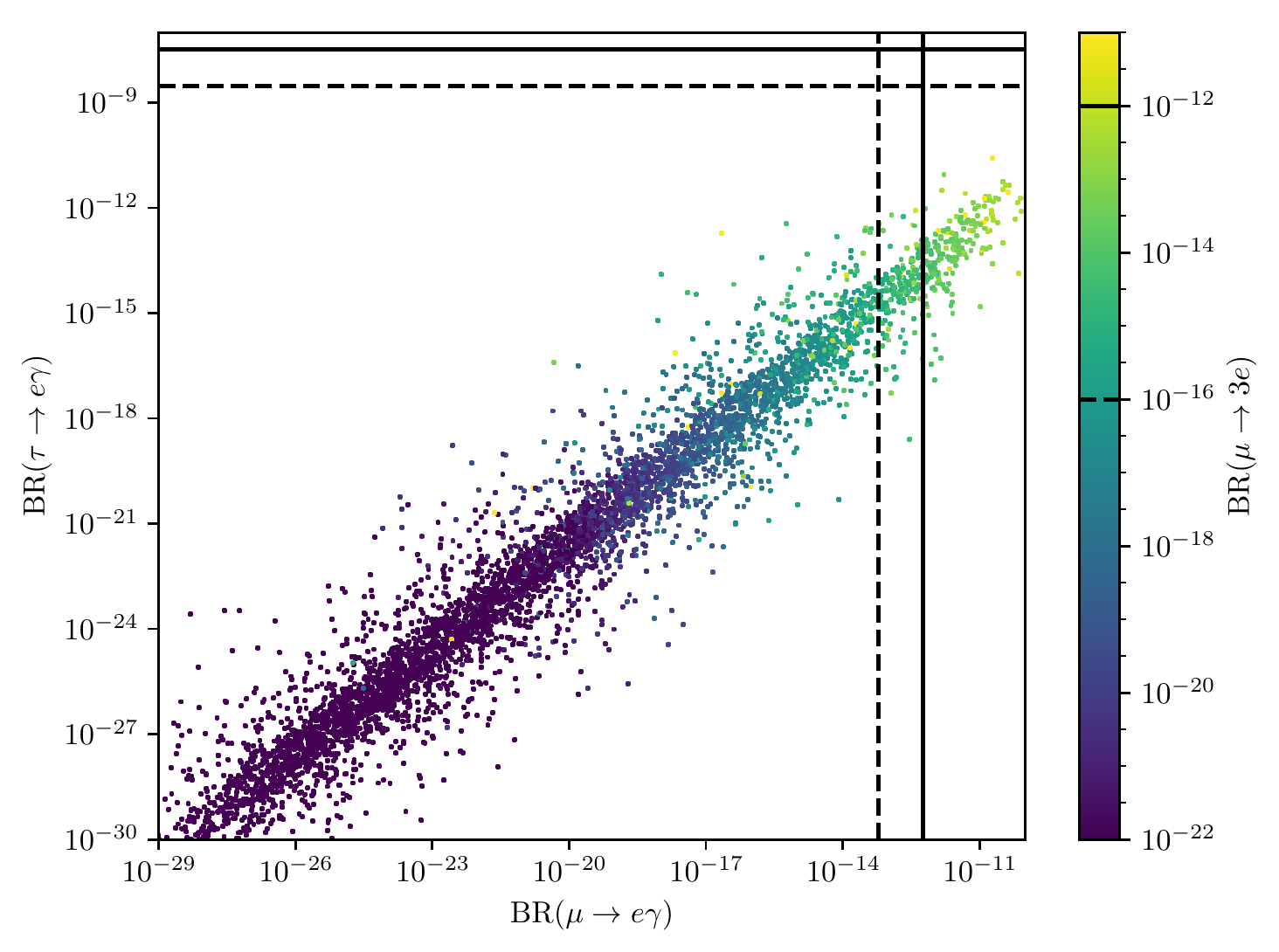}
 \caption{\underline{Left:} The spin-independent direct detection cross section as a function
		of the DM mass for singlet--doublet fermion DM. The colours show the branching
		ratios for the LFV process $\mu \to e \gamma$. Also shown are the LEP limits on
		light neutral and charged particles (shaded areas), current (full lines) and
		future (dashed lines) exclusion limits from XENON1T \cite{Aprile:2018dbl}
		and XENONnT \cite{Aprile:2015uzo}.
   \underline{Right:} Correlations of the branching ratios for the LFV processes
		$\mu \to e \gamma$, $\mu \to 3e$ and $\tau \to e \gamma$ for viable models
		with singlet--doublet fermion DM. Also shown are current (full lines)
		and future (dashed lines) exclusion limits
		\cite{Adam:2013mnn,Baldini:2013ke,Bellgardt:1987du,Blondel:2013ia,Aubert:2009ag,Aushev:2010bq}. \label{fig:2}}
\end{figure}

About one third of all models with the observed neutrino masses and mixings feature singlet--doublet fermion DM, but only
a fraction of order 0.02 \% yield the correct DM relic density
and Higgs mass.
These models are shown in Fig.\ \ref{fig:2} (left) as a function of the DM
mass, together with their spin-independent direct detection cross
section and the branching ratio for the usually most sensitive LFV process
$\mu \to e \gamma$. Other important LFV processes are shown in Fig.\ \ref{fig:2} (right).
The models accumulating at a DM mass of about 1 TeV
feature mostly doublet fermions, whereas lighter fermionic DM is generally
a superposition of singlet and doublet.
A large doublet component below $m_Z/2$ (dark shaded area) is
excluded by the fact that the LEP measurement of the invisible $Z$ boson decay
width is consistent with three generations of active neutrinos.
Furthermore, the accompanying, only slightly
heavier charged fermions are excluded below 102 GeV by largely
model-independent searches with the OPAL detector at LEP (light shaded area).
The LHC limits for heavy long-lived charged particles
from ATLAS and CMS reach currently up to 440 GeV and 490 GeV, respectively, but
are more model-dependent. The spin-independent direct
detection cross section is compared to the current XENON1T exclusion limit
(full line) \cite{Aprile:2018dbl} and the expectation for 20 ton-years with
the XENONnT experiment (dashed line) \cite{Aprile:2015uzo}, which was
extrapolated linearly above 1 TeV. XENON1T excludes most of the
models with small scalar-fermion couplings $\lambda_6$ and therefore also
little LFV. These models are therefore similar to those in the pure
singlet--doublet fermion DM
model. The combination with the scalar sector opens up a considerable
parameter space of leptophilic DM, i.\,e.\ with nuclear recoil cross sections
way below even the expected XENONnT sensitivity. Interestingly, one observes
a strong complementarity with LFV experiments, which already probe the models
with the smallest spin-independent direct detection cross section
\cite{Adam:2013mnn}.

\section{Scalar DM}

\begin{figure}
 \includegraphics[width=0.49\textwidth]{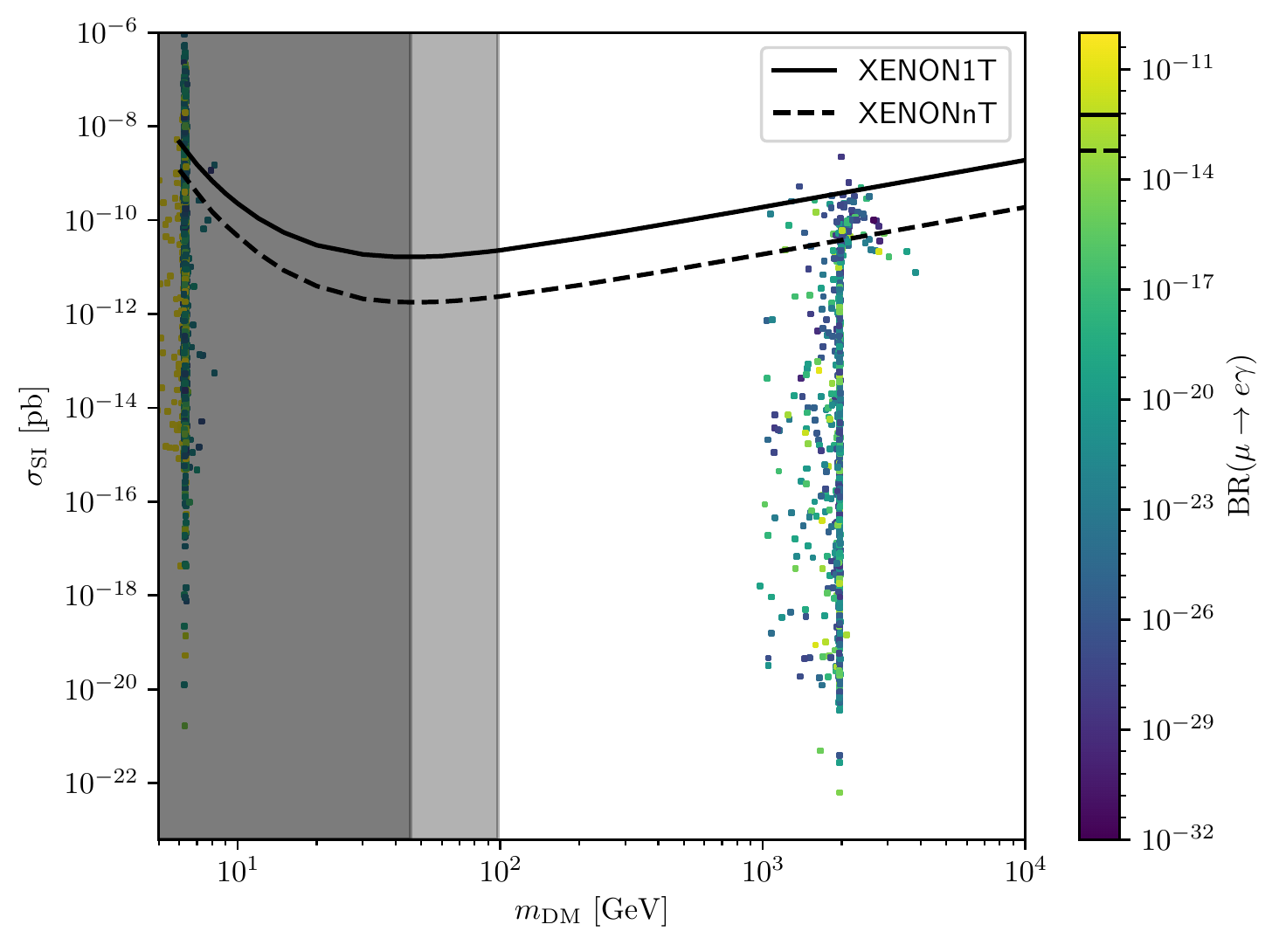}
 \includegraphics[width=0.49\textwidth]{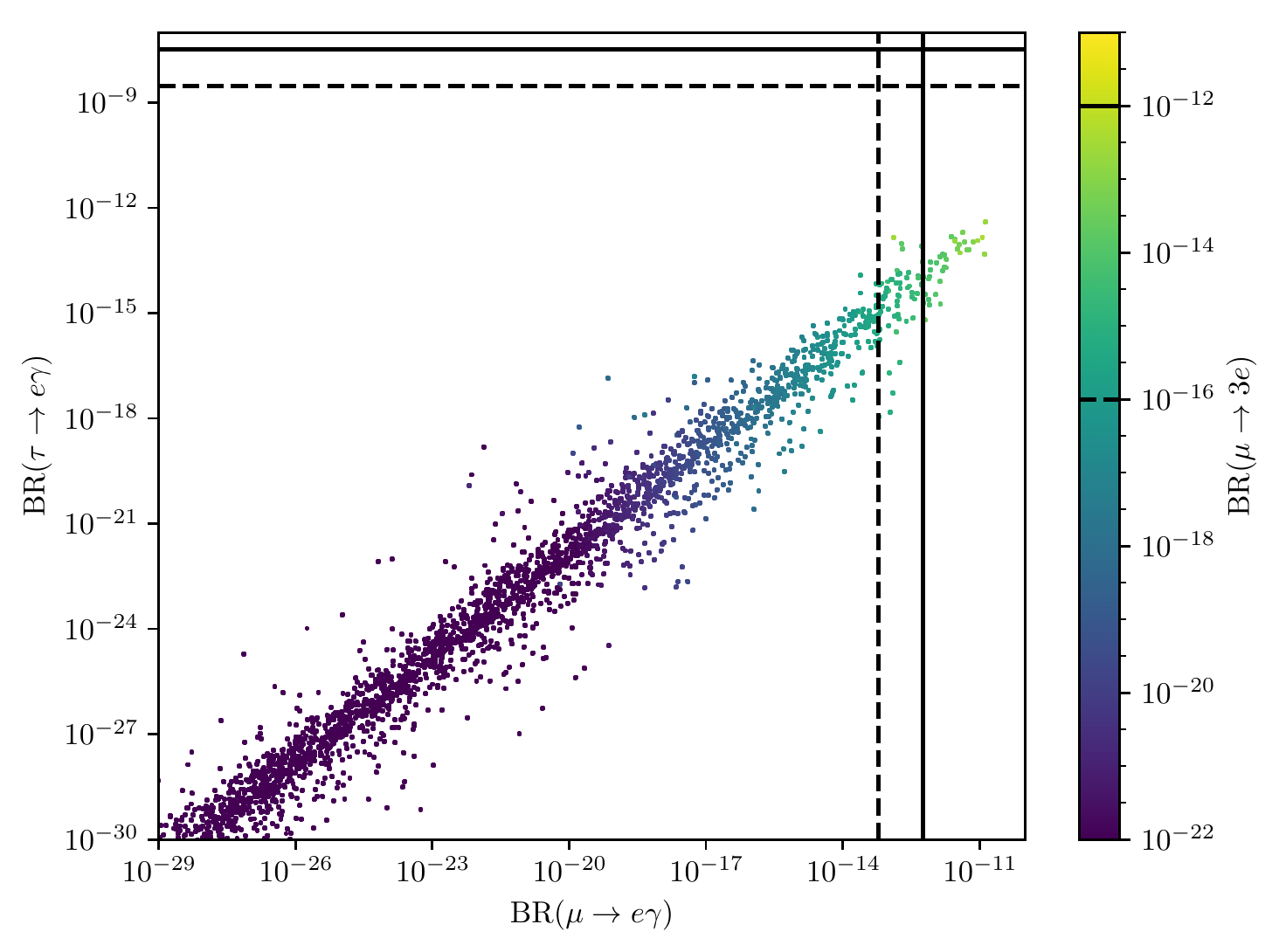}
 \caption{\underline{Left:} The spin-independent direct detection cross section as a function
		of the DM mass for triplet scalar DM. The colours show the branching
		ratios for the LFV process $\mu \to e \gamma$. Also shown are the LEP limits on
		light neutral and charged particles (shaded areas), current (full lines) and
		future (dashed lines) exclusion limits from XENON1T \cite{Aprile:2018dbl}
		and XENONnT \cite{Aprile:2015uzo}.
 \underline{Right:} Correlations of the branching ratios for the LFV processes
		$\mu \to e \gamma$, $\mu \to 3e$ and $\tau \to e \gamma$ for viable models
		with triplet scalar DM. Also shown are current (full lines)
		and future (dashed lines) exclusion limits \cite{Adam:2013mnn,Baldini:2013ke,Bellgardt:1987du,Blondel:2013ia,Aubert:2009ag,Aushev:2010bq}. \label{fig:3}}
\end{figure}

About two thirds of all models with the observed neutrino masses and mixings feature triplet scalar DM, but only
a fraction of order 0.02 \% yield the correct DM relic density and Higgs mass.
These models are shown in Fig.\ \ref{fig:3} (left) as a function of the DM
mass, together with their spin-independent direct detection cross
section and the branching ratio for the LFV process $\mu \to e \gamma$.
 Other important LFV processes are shown in Fig.\ \ref{fig:3} (right).
As for a pure triplet scalar model, we observe an accumulation of points
around a mass of 2 TeV. Many of these models have only very small couplings
$\lambda_6$ to the fermion sector and thus very little LFV. As $\lambda_1$
increases, so must the DM mass beyond 2 TeV to compensate for the stronger
Higgs annihilation. However, most of these models will soon be probed by
XENONnT, and those that will not can soon be excluded by the process
$\mu \to e \gamma$. While the mass region from 1 TeV to 2 TeV with leptophilic
fermion DM, that was opened up by coupling the fermion and scalar sectors, was
already excluded by LFV limits (see above), the corresponding models with
scalar DM are still allowed, but will soon be probed by the process
$\mu \to e \gamma$. Note that there exists in principle also a region of very
light triplet scalar DM of about 6 GeV mass, which is however excluded by
the LEP limits on light non-sterile neutral (dark shaded area) and charged
(light shaded area) particles.

\section{Conclusion}

To summarise, we heave presented a phenomenological study of a radiative seesaw
model with singlet-doublet fermion and triplet scalar DM. For each individual
model, we found that the new XENON1T results doubled the excluded parameter
space for singlet-doublet fermion DM, while for triplet scalar DM the viable
mass was about 2 TeV for small Higgs couplings, but then increased to compensate
for the larger couplings.

The combination of the fermion and scalar sectors required two generations of
scalars for the generation of two non-zero neutrino masses. DM was found to be
fermionic up to 1 TeV, then scalar, and the combination allowed for smaller
masses than the individual models still in agreement with XENON1T data. We
observed a strong complementarity of direct detection and lepton flavour violation
experiments, while the (model-dependent) LHC limits remained relatively weak
at around 440 to 490 GeV.

\end{document}